# Hierarchical Variability Modeling for Software Architectures

Arne Haber*, Holger Rendel*, Bernhard Rumpe*, Ina Schaefer† and Frank van der Linden‡

*Software Engineering
RWTH Aachen University, Germany
http://www.se-rwth.de/

† Institute for Software Systems Engineering
TU Braunschweig, Germany
http://www.tu-braunschweig.de/sse

‡CTO Office, Philips Healthcare
Best, The Netherlands
frank.van.der.linden@philips.com

*Abstract*—Hierarchically decomposed component-based system development reduces design complexity by supporting distribution of work and component reuse. For product line development, the variability of the components to be deployed in different products has to be represented by appropriate means. In this paper, we propose *hierarchical variability modeling* which allows specifying component variability integrated with the component hierarchy and locally to the components. Components can contain variation points determining where components may vary. Associated variants define how this variability can be realized in different component configurations. We present a meta model for hierarchical variability modeling to formalize the conceptual ideas. In order to obtain an implementation of the proposed approach together with tool support, we extend the existing architectural description language *MontiArc* with hierarchical variability modeling. We illustrate the presented approach using an example from the automotive systems domain.

*Keywords*-Component-based System Development; Diverse Systems; Variability Modeling

## I. INTRODUCTION

Component-based system development [1] supports the development of large and complex systems. By decomposing a complex system into a set of components in a hierarchy, design complexity can be reduced. The overall functionality of the system becomes more comprehensible by considering the functionality of the components. The component-based decomposition supports the division of work and increases the reuse potential as new applications can be constructed from existing components. A component-based system structure can be exploited for distributed development and component reuse between different product members of the product line.

An important ingredient of product line engineering is a variability model, capturing the diversity of the products contained in the product line. Variability modeling approaches can be classified whether they are concerned with the problem space or with the solution space [2] of the systems under development. The problem space defines the scope of the product line and its member products. Usually, problem space variability is captured in terms of product features. The valid feature combinations can be described by feature models [3] and correspond to the valid member products. The solution space focusses on the reusable artifacts that are customized and composed to realize the actual products. Most approaches representing solution space variability, e.g. [4], [5], [6], deliver a monolithic variability model that specifies the variability of the whole system. Variability specifications are scattered over the entire system architecture and across different hierarchical layers such that for distributed development the variability model has to be kept synchronized for all developers which is time consuming and error-prone. The orthogonal variability model (OVM) [7] provides a low-level decomposed variability model, but is kept separate from the component hierarchy. However, in order to facilitate the development of variable components in a distributed way, component variability has to be represented inline with the component hierarchy. Furthermore, it has to be defined locally to the components. Locality in this context means that the variability of a component is defined either on the hierarchical level of the component or at most one level above or below. Thus, by only considering the component itself all necessary information on the component variability and configuration is available such that each component can be developed independently of the others.

In this paper, we propose *hierarchical variability modeling* as a variability modeling approach for solution space variability that supports the component-based development of diverse systems, such as product lines. *Hierarchical variability* integrates the description of component variability and the representation of the component hierarchy in one model. The variability of components is specified by considering only the components themselves. A variable component contains *variation points* representing *what*

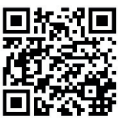

*can vary* in the component. A variation point has a set of associated *variants* that describe *how the component varies* [7]. The variants can be described again by variable components or selected variants of variable components such that component variability is specified together with the component hierarchy.

In order to provide a concrete language for hierarchical variability modeling of software architectures, we extend the existing architecture description language (ADL) MontiArc [8], [9] by variation points as specific architectural modeling elements to represent variable components. Variation points can be realized by associated variants that contain a set of architectural modeling elements to be added to the component if the respective variant is selected. The resulting language for hierarchical variability MontiArc$^{HV}$ supports the modular modeling of architectural variability within the component hierarchy. Component variability is encapsulated in the single components such that system evolution can be handled by local component modifications. In particular, *hierarchical variability modeling supports evolution by adding or removing variants*. The compositional representation of variability, further, provides means for compositional validation and verification of the resulting architecture.

Hierarchical variability modeling, as presented in this paper, extends earlier work on hierarchical variability [10], [11]. The Koala component model [12], [13], [14], [15] was a first attempt to represent variability inline with the component hierarchy by explicit linguistic constructs inside the components. However, Koala is a component model only available during implementation, compilation, and run-time. Hence, the language concepts for variability representation are tailored only to the implementation level. The hierarchical variability modeling language MontiArc$^{HV}$ described in this paper extends Koala and aims at providing support for hierarchical variability modeling in component-based system development in the early development phases, in particular during architectural design.

This paper is structured as follows. In Section II, the need for hierarchical variability modeling in component-based development of diverse systems is motivated. Section III presents component-based system development with MontiArc. In Section IV, the meta model for hierarchical variability modeling is discussed. Its usage is illustrated using an example from the automotive domain in Section V. The ADL for hierarchical variability MontiArc$^{HV}$ is described in Section VI. Section VII outlines a modeling procedure for hierarchical variability modeling. Related approaches are discussed in Section VIII. The paper is summarized in Section IX.

## II. SYSTEM DIVERSITY

Diversity is prevalent in modern software systems. Systems exist in many different variants in order to adapt to their application context. Diversity stems from many sources: some are hardware-related, others are cultural, and further sources are related to choices in quality attributes of the system (resource consumption, usability, security, ...). Each source may lead to several variations of the system. Combined, they give rise to an even larger number of possible system variants.

Product line engineering [7] aims at developing a set of diverse systems with well-defined commonality and variability by managed reuse. It benefits from component-based development approaches. Decomposing the systems to be developed into components supports the distribution of work and the communication between different development teams. Additionally, the component-based system structure facilitates reusing similar components in different products.

If a product line is created in a component-based approach, the diversity of the products is captured by the diversity of the used components. A component may occur in different variants in the system. These variants are constituted by selecting concrete variants for the contained variable subcomponents. In general, the diversity of a component can be described in two ways:

1) The *variability* describes the possible variants of a variable component. The variability of a component can be restricted, e.g., by defining constraints on the variants of the contained variable sub-components.
2) The *configuration* of a component describes which concrete variant of the component is chosen in a system in which this component occurs. The configuration information is, for instance, important to determine which systems need to be updated in case a new version of the component variant becomes available.

In order to support the component-based development of diverse systems, a variability model for the solution space is required that integrates the representation of component variability with the component hierarchy. This modelling approach for hierarchical variability has to satisfy the following requirements:

1) Component variability and component hierarchy need to be treated uniformly in one model.
2) Variability must be specified locally to the components. Constraints on variant selection should not go across hierarchical boundaries in order to facilitate the distributed development of variable components.
3) The variability model should allow focussing on the common architecture of all system variants, on the component variability and on the used component configurations, since at any point in time during system development, maintenance and evolution, it is important to know which variants can be supported and which configurations are actually used.
4) Design/configuration decision on a high level maps

```
1  component LockControlUnit {
2    autoconnect port;
3  
4    port
5      in OpenCloseRequest,
6      out LockStatus;
7  
8    component LockActuator;
9    component LockController;
10 
11   connect LockActuator.cmdSucceeded ->
12       LockController.lockSuccess;
13 }
```

Listing 1. Structural component `LockControlUnit` in MontiArc syntax

down to variant selection on components on a low level of the hierarchy.

## III. COMPONENT-BASED SYSTEM DEVELOPMENT

Component-based system development [1] aims at developing large and complex systems by decomposing the overall system into a set of components. Components can be themselves decomposed into subcomponents providing a hierarchical system structure. A component-based system can be described using an architecture description language (ADL). An ADL describes the components, their subcomponent structure, the component interfaces, which are specified by ports, and the associated communication connections.

We base our hierarchical variability modeling approach for software architectures on the existing textual ADL *MontiArc* [8]. MontiArc is designed for modeling distributed component-based information flow architectures in which communication is based on asynchronous messages. We decided to use MontiArc as a base language for our hierarchical variability approach, and not similar languages like Acme [16] or xADL [17], because we require the extensibility of human readable concrete syntax and language tool support. MontiArc is developed using the DSL framework MontiCore [18] that supports language reuse on concrete and abstract level of textual DSLs such that the required extensibility is achieved.

Architectural components in MontiArc are units of computation or storage defining their computational commitments via interfaces [19]. Theses interfaces are the only interaction points of components to provide clear concepts of interaction between entities of computation [20]. Logical architectures modeled by MontiArc can be realized in software or in hardware.

An example of a MontiArc component is given in Listing 1 that shows the architecture of a lock control unit that controls locking and unlocking of a car-door. It receives a request, triggers the contained subcomponents and emits the status of the controlled lock. Based on this example the main modeling-elements of MontiArc are introduced. A component definition starts with the keyword `component` followed by the components name (cf. l. 1) and its interior surrounded by curly braces. As components are organized in packages they can be uniquely identified by a qualified name. Implementation of a component is either given by decomposition to subcomponents (*structural component*) or a Java implementation or state chart realizing the components behavior (*atomic components*). A subcomponent starts with the keyword `component` followed by a component type and an optional name (cf. ll. 8-9). This way a subcomponent is an instance of it's component-type.

Component interface definitions are given by ports (cf. ll. 4-6). A port always has a direction - `in` for incoming and `out` for outgoing ports - and a data type. In MontiArc, explicit naming of ports as well as subcomponents is optional as long as their type is unique in the current component definition. If implicit naming is used, a port respectively a subcomponent is named after its type. Explicit names are given after the type of a subcomponent or port.

To ease modeling, MontiArc offers an implicit mechanism to create communication connections. The `autoconnect port` statement (cf. l. 2) implicitly connects ports with the same unique name and a compatible type. If it is parametrized with `type` instead of `port` implicit connections between ports with the same unique type are created disregarding the ports' names. If it is not possible to create all connections automatically, as uniquely identifying names may not be always given, explicit connections can be created using the `connect` statement connecting one source port with one or more target ports. If a target or source port belongs to a subcomponent, it is qualified with the subcomponent's name. Implicit connectors can always be redefined by explicit connector definitions. This way the connector in ll. 11f connects the outgoing port `cmdSucceeded` of subcomponent `LockActuator` with the incoming port `lockSuccess` of subcomponent `LockController`.

## IV. HIERARCHICAL VARIABILITY MODELING

*Hierarchical variability* describes component variability and component hierarchy uniformly in one model. We distinguish three categories of modeling elements in hierarchical variability modeling, as shown in the meta model in Fig. 1: *commonality*, *variability* and *configuration*.

*Commonality* captures the common architecture of all systems. These common architectural modeling elements are defined by the ADL MontiArc (cf. Section III) that serves as basis for the presented approach. A `Component` is implemented by a set of architecture elements (`ArcElement`). An `ArcElement` can be a component, a subcomponent, a port, or a connector.

*Variability* of components is defined by *variation points*. Following the terminology of Pohl et al. [7], a variation point describes what can vary in a component. In order to express this variability, the MontiArc component definition is extended by variation points (`VariationPoint`) that can be used as additional `ArcElement`. Components containing variation points are referred to as *variable components*.

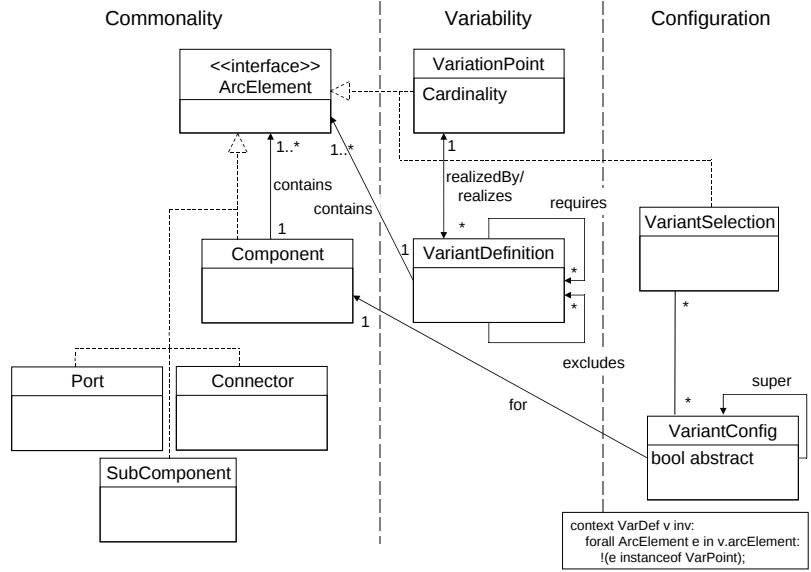

Figure 1. Meta model for Hierarchical Variability Modeling

Each variation point has a set of associated variants that specify how the variation point can be realized. Following [7], *variants* specify how the component can vary. Variants are defined by the concept `VariantDefinition` and are connected to exactly one variation point. A variant contains several `ArcElements` which are added to the component definition if the variant is selected to realize the corresponding variation point. A variation point has a cardinality which defines how many variants can be or have to be selected to realize the variation point. Using the cardinality, optional variants ([0..1]), required variants ([1..1]), or arbitary multiplicities for the variant selection can be expressed. The selection of a variant at one variation point may require or exclude the selection of a specific variant at another variation point of the component. This can be expressed by `requires` and `excludes` variability constraints between variants on the same level of hierarchy. Variants may not contain variation points themselves, which is expressed by the OCL constraint in Fig. 1, in order to allow for the local specification of component variability. Instead, a variant may contain variable subcomponents that encapsulate the variability of a set of component variants sharing a common architecture.

The *configuration* describes the selection of component variants and the configuration of the actual products defined by the hierarchical variability model. For the definition of a component, it might be necessary to restrict the variability of contained variable subcomponents and to select a particular subcomponent variant. A variant selection (`VariantSelection`) allows configuring the variation points of inner subcomponents by selecting the desired associated variants. A specific component variant that is the result of a variant selection can also be used as `ArcElement` in other component definitions. With variant selections in the component hierarchy, only variation points of direct subcomponents can be configured in order to maintain the locality of the component definition.

A complete product configuration (`VariantConfig`) consists of variant selections for all variable components contained in the system that are not selected in the component hierarchy. Only in a product configuration using `VariantConfig`, it is possible to configure variation points at arbitrary positions in the component hierarchy, since locality of the specification is only relevant for components that should be reused, but not for the actual products. In hierarchical variability modeling, it is possible to provide partial system configurations by leaving some variation points underspecfied, for instance, to define a common system platform where only configurations of variable subcomponents at lower hierarchical levels might be different. A partial system configuration can be provided allowing abstract `VariantConfigs`. Abstract configurations can be extended iteratively by adding variant selections for the underspecfed variation points. The reconfiguration of variation points by replacing previously selected variants is not allowed since this would invalidate the previously defined system configuration. Hence, the relationship between abstract partial configurations and complete configurations can be described by inheritance.

Hierarchical variability modeling satisfies the requirements formulated in Section II: It allows expressing component hierarchy and component variability uniformly in one model by considering variable components and variant selections likewise as architectural modeling elements. Com-ponent variability is specified locally by allowing variability constraints and variant selections only on the same or adjacent hierarchical levels. The supported component variants can be derived from the variable component definitions by considering all possible variant selections under the given variability constraints. The used component variants are determined by the specified variant selections.

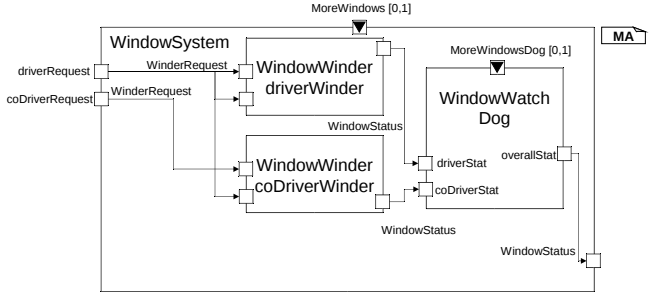

Figure 2. Component WindowSystem

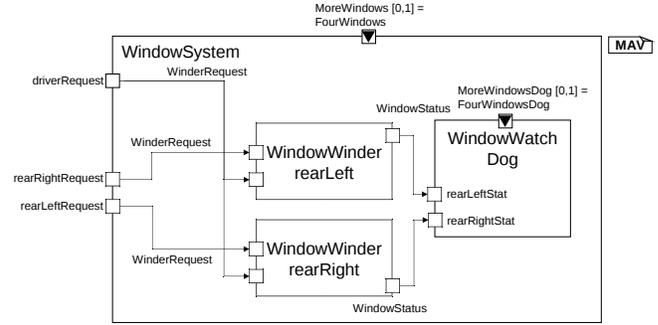

Figure 3. Variant FourWindows

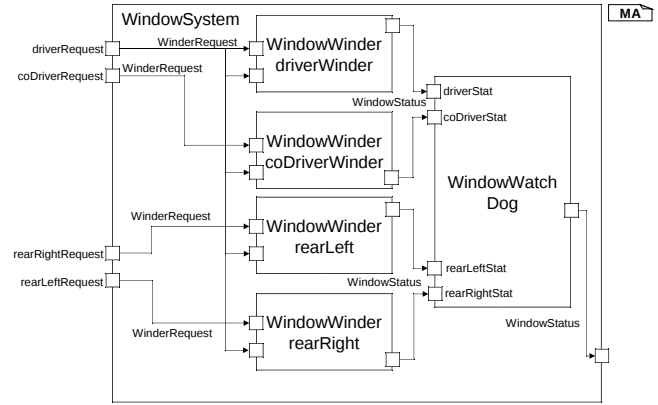

Figure 4. Resulting system for configuration FourWindowSytem

## V. Illustrative Example for Hierarchical Variability

As an example for hierarchical variability, we consider a car with an electric window control system that manages requests for electric windows from different switches. This system can be realized in software or in hardware which demonstrates that this approach is not restricted to one of these domains. The driver of the car is allowed to control all windows. In addition, each window can be controlled by the corresponding passenger. The system is modeled as a component `WindowSystem` depicted in Fig. 2. The subcomponents are a `WindowWinder` component for the driver window and a `WindowWinder` component for the co-driver window. As input, every `WindowWinder` component receives the request by the driver and the request by the respective passenger. It gives priority to the `DriverRequest`, in case the incoming requests differ. The outputs of all `WindowWinder` components are merged in a component `WindowWatchDog`. The `WindowWatchDog` component monitors the status of the windows and produces a signal containing the actual status of all windows. This status is evaluated by the window motors to trigger an action (e.g. moving down a specific window).

The `WindowSystem` component has a variation point `MoreWindows` marked by the triangle on the top edge of the component in Fig. 2. The variation point enables the optional addition of architectural elements to support a system with more than two windows. The subcomponent `WindowWatchDog` also has a variation point `MoreWindowsDog` allowing to adjust this component for dealing with more than two windows.

A possible variant `FourWindows` for variation point `moreWindows` is given in Fig. 3. This variant adds two `WindowWinder` components for the rear windows, the required ports, and connections. Additionally, the subcomponent `WindowWatchDog` needs to be changed to deal with for the additional two windows. This is done by selecting the variant `FourWindowsDog` for variation point `MoreWindowsDog` of `WindowWatchDog` component.

An example of a configured system that is defined by the hierarchical variability model is depicted in Fig. 4. In this configuration, the variation point `MoreWindows` is configured with the variant `FourWindows` which automatically configures the variation point of the contained subcomponent `MoreWindowsDog`.

The textual representation of the `WindowSystem` component in our architectural description language with hierarchical variability MontiArc$^{HV}$ is given in Listing 2. The listing contains the MontiArc defintion of the common architectural elements, including ports (cf. ll. 3-6), references to the `WindowWinder` subcomponents (cf. ll. 8-9), the subcomponent `WindowWatchDog` that is defined inside the component (cf. ll. 11-18), and the required connections (cf. ll. 20-31). Additionally, the variation points for the `WindowWatchDog` component and for the `WindowSystem` component are specified. (cf. ll. 17, 33). The cardinality of both ([0..1]) marks them as optional variation points.

Listing 3 defines the variant `FourWindows` for the variation point `MoreWindows` in Listing 2 (cf. l. 33). The

```
component WindowSystem {

  port
    in WinderRequest driverRequest,
    in WinderRequest coDriverRequest,
    out WindowStatus;

  component WindowWinder driverWinder,
      coDriverWinder;

  component WindowWatchDog {
    port
      in WindowStatus driverStat,
      in WindowStatus coDriverStat,
      out WindowStatus overallStat;

    variationPoint: MoreWindowsDog [0..1];
  }

  connect driverRequest ->
      driverWinder.driverRequest,
      driverWinder.passengerRequest,
      coDriverWinder.driverRequest;
  connect coDriverRequest ->
      coDriverWinder.passengerRequest;
  connect driverWinder.WindowStatus ->
      WindowWatchDog.driverStat;
  connect coDriverWinder.WindowStatus ->
      WindowWatchDog.coDriverStat;
  connect WindowWatchDog.overallStatus ->
      WindowStatus;

  variationPoint: MoreWindows [0..1];
}
```

Listing 2. Component WindowSystem

```
variant FourWindows realizes
    WindowSystem.MoreWindows {
  port
    in WinderRequest rearLeftRequest,
    in WinderRequest rearRightRequest;

  component WindowWinder rearLeft, rearRight;

  connect driverRequest ->
      rearLeft.driverRequest,
      rearRight.driverRequest;
  connect rearLeftRequest ->
      rearLeft.passengerRequest;
  connect rearRightRequest ->
      rearRight.passengerRequest;
  connect rearLeft.WindowStatus ->
      WindowWatchDog.rearLeftStat;
  connect rearRight.WindowStatus ->
      WindowWatchDog.rearRightStat;

  WindowSystem.WindowWatchDog.MoreWindowsDog
      realizedBy FourWindowsDog;
}
```

Listing 3. Variant FourWindows

```
variantConfig FourWindowSystem for
    WindowSystem {
  WindowSystem.MoreWindows
      realizedBy FourWindows;
}
```

Listing 4. Configuration FourWindowsSystem

variant adds ports for the input of the rear windows (cf. ll. 3-5), the `WindowWinder` subcomponents for the rear windows (cf. l. 7), and the required additional connections. Since the `WindowWatchDog` component must be adjusted if this variant is selected, its variation point is configured by selecting the variant `FourWindowsDog` (cf. ll. 21-22).

An example system configuration in MontiArc$^{HV}$ is given in Listing 4. To obtain a valid system architecture for a system with four windows, the top level variation point `WindowSystem.MoreWindows` must be configured by selecting the variant `FourWindows` (cf. ll. 3-4). This selection also configures the variation point of the variable subcomponent `WindowWatchDog`

The variable component `WindowSystem` can now be reused to model a car with an electric window control system and an electric locking system. Thus, the `Car` component contains two variation points to capture the variability of the electric locking and the window control systems. The first variation point, called `LockController`, allows the addition of an eletric locking system component `LockControlUnit` as described in Sect. III by selecting the associated variant `FourDoorsLock`. The second vari-ation point, called `WindowController`, is used to add a specific variant of the electric window control component `WindowSystem`.

The variant `FourWindowVehicle` is one possibility to realize the variation point `WindowController`. A four window vehicle in this example always requires the presence of the electric locking system. Hence, there is a variability constraint between the variants that are associated to the variation points `LockController` and `WindowController`. This dependency can be descried as depicted in Listing 5 (cf. l. 3) using a `requires` clause. The variant `FourWindowVehicle` also adds a `WindowSystem` component (cf. l. 5) and configures its variation point (cf. l. 7-8) such that the variant `FourWindows` is selected.

```
variant FourWindowVehicle realizes
    Car.WindowController
    requires LockController.FourDoorsLock {

  component WindowSystem;

  WindowSystem.MoreWindows realizedBy
      FourWindows;
}
```

Listing 5. Variant FourWindowVehicle

```
1  grammar HierVarArc
2       extends mc.umlp.arc.MontiArc {
3    VariationPoint implements ArcElement =
4      "variationPoint" ":" Name
5      Cardinality? ";";
6  }
```

Listing 6. MontiCore grammar that adds variation points to component definitions

## VI. AN ADL FOR HIERARCHICAL VARIABILITY

To provide tool support for hierarchical variability modeling, we implemented the language MontiArc$^{HV}$ using the MontiCore framework [18]. Since we want to add hierarchical variability modeling to the existing ADL MontiArc, we extend the MontiArc [8] grammar using MontiCore's language inheritance mechanism. MontiCore is used to specify the standard architectural modeling elements. MontiArc$^{HV}$ extends MontiArc with the concepts shown in Listing 6 in order to define the part of MontiArc$^{HV}$ describing variable components with variation points. The production `VariationPoint` implements the interface `ArcElement` that is given in the MontiArc super-grammar `mc.umlp.arc.MontiArc`. A component described in this language can contain variation points in addition to MontiArc ArcElements. The specification of a variation point starts with the keyword `variationPoint:` followed by its name and an optional (?) cardinality. If no cardinality is given we assume that the variation point is realized by exactly one variant. The part of the MontiArc$^{HV}$ grammar for the the specification of variants is given in Listing 7. It also extends the MontiArc language and inherits the MontiArc architectural modeling elements. The starting rule `VariantFile` allows processing variant definitions (`VariantDefinition`) or component configurations (`VariantConfig`).

A variant definition `VariantDefinition` (cf. ll. 6-11) starts with the keyword `variant` followed by its name. Variant definitions in MontiArc$^{HV}$ can be parametrized by optional variant parameters (cf. ll. 15-17). Using these parameters, ports or subcomponents that are added by the variant get a parametric name. Technically, this is implemented by overwriting the production rules for ports and for the name part of subcomponents to add a `variableParameter` to the names separated by ∼ (cf. ll. 27, 30). The parameters are instantiated with concrete values in a variant selection as explained below. Parametrization is, for instance, useful if the same variant should be selected several times, but the added ports or components should have different names.

A variant definition always `realizes` a specific variation point that is referenced by its qualified name. An optional `constraint` is used to declare variability constraints between variants. In the meta model (cf. Fig. 1) and the example (cf. Sect. V), we used `requires` and

```
1  grammar VariantDefinitionDSL
2       extends mc.umlp.arc.MontiArc {
3    VariantFile =
4      (VariantDefinition | VariantConfig);
5
6    VariantDefinition =
7      "variant" Name VariantParameter?
8      "realizes" variationPoint:
9          QualifiedName
10     ("constraint" "(" Constraint ")")?
11     body:ArcComponentBody;
12
13   external Constraint;
14
15   VariantParameter =
16     "(" parameters:Name
17     ("," parameters:Name)* ")";
18
19   VariantReference =
20     (variantName:QualifiedName
21     ("(" parameters:Name
22     ("," parameters:Name)* ")" )? );
23
24   ArcPort =
25     Stereotype?
26     (incoming:["in"] | outgoing:["out"])
27     Type (variableParameter:Name "˜")?
28     Name?;
29
30   ArcReferenceInstance =
31     (variableParameter:Name "˜")? Name;
32
33   VariantConfig =
34     ...
35 }
```

Listing 7. MontiCore grammar for the definition of variants

`excludes` constraints. However, the external production `Constraint` can be instantiated with a more general constraint language (e.g., OCL) using MontiCore's language embedding mechanism to express more complex variability constraints. The body of a variant definition is given by an `ArcComponentBody` which it may contain arbitrary many `ArcElements` including variant selections. A component configuration `VariantConfig` selects concrete variants for the variation points contained in a variable component. This part of the grammar is not shown in Listing 7.

Based on the grammars for MontiArc$^{HV}$ depicted in Listings 6 and 7, MontiCore generates an infrastructure to process concrete hierarchical variability models in MontiArc$^{HV}$ including a lexer, a parser and and data structures for representing the abstract syntax tree. However, not all MontiArc$^{HV}$ models that are syntactically correct are meaningful. To find faulty models, context conditions have to be defined and implemented. For MontiArc$^{HV}$, the following set of context conditions extends the context conditions that have to hold for standard MontiArc (cf. [8]) models:

1) Non-abstract variant configurations have to configure all variation points in the referenced architecture, either by defining variant selections themselves or by extending a super-configuration.
2) The component referred to in a variant configuration has to exist, and contain all variation points that are referenced.
3) The super-configuration of a configuration has to exist.
4) The variation point that is realized by a variant definition has to exist.
5) Referenced variation point and the referenced variants in variant selections have to exist.
6) The number of actual parameters that are provided by a variant selection has to match the number of formal parameters of the corresponding variant definition.

## VII. Modeling with MontiArc$^{HV}$

MontiArc, as the basis ADL for MontiArc$^{HV}$, supports bottom-up as well as top-down modeling approaches, such as the Attribute Driven Design Method [21]. Because MontiArc$^{HV}$ is an extension of MontiArc, this holds for MontiArc$^{HV}$ too. In this section, we sketch a top-down method for modeling a variable component structure with MontiArc$^{HV}$ that is based on the three modeling categories: commonality, variability, and configuration.

First of all, the common parts of a system architecture have to be defined. The system is split into common subsystems (subcomponents). Communication between these is modeled by connecting component interfaces that are given by ports. The contained subcomponents are decomposed themselves into common architectural elements. All common elements of the modeled system, i.e., common parts of variable components, subcomponents and non-variable components, will later be part of every defined system.

As a second step, variability is introduced in the model. Therefore, variation points are added to components to define explicitly where components may vary. Then, variants that can be selected to realize the variation points are modeled. The addition of variability to a component may imply that variability has to be introduced to the contained subcomponents as well. As an example, consider a variant $V_1$ that adds a port $P_1$ to a component $C$. The port $P_1$ should be connected to the contained subcomponent $SC$. Hence, a variation point $VP_{SC}$ has to be added to the subcomponent $SC$, and an associated variant $V_2$ has to be defined that adds the corresponding port to subcomponent $SC$ in order to provide the desired connection. Additionally, in variant $V_1$, a variant selection of the variation point $VP_{SC}$ has to be defined such that the variant $V_2$ in subcomponent $CS$ is selected whenever variant $V_1$ is selected.

The third step of the modeling method is to define system configurations by selecting variants for variation points respecting their cardinality. A concrete configuration corresponds to a product, while abstract configurations correspond to a common platform. The open variation points of abstract configurations are configured later to describe concrete products. The above steps may be iterated, until a desired degree of decomposition is achieved.

## VIII. Related Approaches

The reusable artifacts of a component-based software product line have to be organized such that an efficient product derivation is faciliated [22]. Different alternatives are discussed in [23], [24], such as structuring the variability models according to the organizational structures, specific business needs or following the architectural components. However, in these works, a designated variability model for diverse component-based systems is not defined. Existing approaches to represent solution space variability models of component-based systems can be classified in two main directions [25]: annotative (or negative) and compositional (or positive) modeling approaches.

Annotative approaches consider one model (that is usually non-hierarchical) representing all products of the product line. Variant annotations, e.g., using UML stereotypes [4], [5] or presence conditions [6], define which parts of the model have to be removed to derive a concrete product model. The orthogonal variability model (OVM) [7] models the variability of product line artefacts in a variability model that is separated from the artifact model. Links from the variability model to the artifact model take the place of annotations and determine which model parts are removed for certain product variants. In [26], the variability modeling language (VML) is proposed that specializes the ideas of OVM for architectural models.

Compositional approaches associate model fragments with product features that are composed for a particular feature configuration. In [27], [25], [28], models are constructed by aspect-oriented composition. Feature-oriented model-driven development (FOMDD) [29] combines feature-oriented programming (FOP) with model-driven engineering. In [30], model fragments are merged in order to provide the variability model of a product line. Apel et al. [31] apply model superposition to compose model fragments. Model superimposition considers models with a hierarchical structure that is preserved when models are composed. But the approach presented in [31] does not focus on supporting component-based software development.

Apart from positive and negative variability representations, model transformations are used for capturing product variability. The common variability language (CVF) [32] represents the variability of a base model by rules describing how modeling elements of the base model have to be substituted in order to obtain a particular product model. In [33], graph transformation rules capture the variability of a single kernel model comprising all commonality. In [34], architectural variability is represented by change sets containing additions and removals of components and component connections that are applied to a base line architecture. Delta modeling [35], [36] is a modular approach to represent system variability via transformations. A diverse set of systems is represented by a designated core system and a set of system deltas explicitly specifying changes to the core system in order to obtain other system variants. Delta modeling can be used for component-based systems [37], but an integration of delta modeling with the component hierarchy is not yet considered.

Plastic partial components [38] provide a means to model component variability inside the components by extending partially defined components with variation points and associated variants. Variants can be cross- or non-cross-cutting architectural concerns that are composed with the common component architecture by weaving mechanisms that have to be specified by the component designer. However, variants cannot contain variable components such that component-variability and -hierarchy are not fully integrated.

The Koala component model [13], [39] is a first approach aiming at hierarchical variability modeling. In Koala, the variability of a component is described by the variability of its sub-components. The selection between different sub-component variants is realized by *switches* that are used as designated components. Via explicit *diversity interfaces*, information about selected variants is communicated between sub- and super-components in order to configure the switches to select a specific subcomponent variant. Diversity interfaces and switches in Koala can be understood as concrete language constructs targeted at the implementation level to express variation points and associated variants. The approach presented in this paper abstracts from these concrete language constructs. Instead, it provides a general approach that integrates component variability and component hierarchy to foster component-based development of diverse systems during architectural design.

Other textual ADLs that may be extended with hierarchical variablity modeling concepts are Acme [16] or xADL [17]. An extension of Acme can be achieved using its property mechanism. A variation point may be embedded in a property and, hence, modeled as a plain string. That, however, is an error prone approach, as these strings has to be interpreted manually. Syntax highlighting or further modeling support for variation points cannot be provided by Acme. xADL can be extended by defining new XML schemes. As human readability of XML files is poor, xADL does not match our extensibility requirements, too.

## IX. CONCLUSION

Hierarchical variability modeling integrates the representation of the component variability with the component hierarchy. Component variability is expressed locally within components, such that distributed development of variable components is facilitated. Using hierarchical variability modeling, it is possible to determine the component variants that are supported and the variants that are currently used at all times during system development and maintenance.

For future work, we aim at improving our tool support for MontiArc$^{HV}$ and at evaluating the modeling concepts using industrial-scale case examples of diverse component-based systems. In order to examine the advantages of hierarchical variability modeling, we plan an detailed conceptual and empirical comparison with other variability modeling approaches for component-based systems, such as [40] or [37]. Future work will also evaluate how evolution can be supported by hierarchical variability modeling via the addition and removal of variants. Currently, MontiArc$^{HV}$ supports only the addition of architectural modeling elements to an existing component. In the future, MontiArc$^{HV}$ will be extended by more invasive composition techniques, such as model superimposition [31], to facilitate more expressive components modifications by selected variants. MontiArc$^{HV}$ is designed to model solution space variability of system components. In order to provide a connection to the problem space variability that is usually described by feature models, we plan to investigate how the configuration of variation points by variant selections relates to the selection of problem space features.